\documentclass[journal=jacsat,manuscript=article]{achemso}
\usepackage{graphicx}
\usepackage{dcolumn}
\usepackage{bm}
\usepackage{amssymb}
\usepackage{multirow}
\usepackage{color}

\newcommand{\1}{\begin{equation}}
\newcommand{\2}{\end{equation}}
\newcommand{\ea}{\begin{eqnarray}} 
\newcommand{\ee}{\end{eqnarray}}
\newcommand{\4}[2]{{\frac{#1}{#2}}}
 
\newcommand{\Sum}[2]{{\sum\limits_{#1}^{#2}}}

\author{Benno Liebchen}
\affiliation{Institut f\"{u}r Theoretische Physik II: Weiche Materie, Heinrich-Heine-Universit\"{a}t D\"{u}sseldorf, D-40225 D\"{u}sseldorf, Germany}
\email{liebchen@hhu.de}
\author{Hartmut L\"owen}
\affiliation{Institut f\"{u}r Theoretische Physik II: Weiche Materie, Heinrich-Heine-Universit\"{a}t D\"{u}sseldorf, D-40225 D\"{u}sseldorf, Germany}
\email{hlowen@hhu.de}

\title[]{Synthetic Chemotaxis and Collective Behavior in Active Matter}

\abbreviations{IR,NMR,UV}

\begin{document}
\textbf{Conspectus:}
The ability to navigate in chemical gradients, called chemotaxis, is crucial for the survival of microorganisms. 
It allows them to find food and to escape from 
toxins. 
Many microorganisms can produce the chemicals to which they respond themselves and use 
chemotaxis for signalling which can be seen as a basic form of communication, allowing 
ensembles of microorganisms to coordinate their behavior. This occurs during processes like embryogenesis, biofilm formation or cellular aggregation. 
For example, Dictyostelium cells use signalling as a survival strategy: 
when starving they produce certain chemicals towards which other cells show taxis. 
This leads to aggregation of the cells resulting in a multicellular aggregate which can sustain long starvation periods. 

Remarkably, the past decade has let to the development of synthetic microswimmers, 
which can self-propel through a solvent, analogously to bacteria and other microorganims. 
The mechanism underlying the self-propulsion of synthetic microswimmers like 
camphor boats, droplet swimmers and in particular autophoretic Janus colloids
involves the production of certain chemicals.
As we will discuss in this Account, the same chemicals (phoretic fields) involved in the self-propulsion of a (Janus) microswimmer also acts on other 
ones and biases their swimming direction towards (or away from) the producing microswimmer. 
Synthetic microswimmers therefore provide 
a synthetic analogue to motile microorganisms interacting by taxis towards (or away from) self-produced chemical fields.

\begin{figure}
\includegraphics[width=0.3\textwidth]{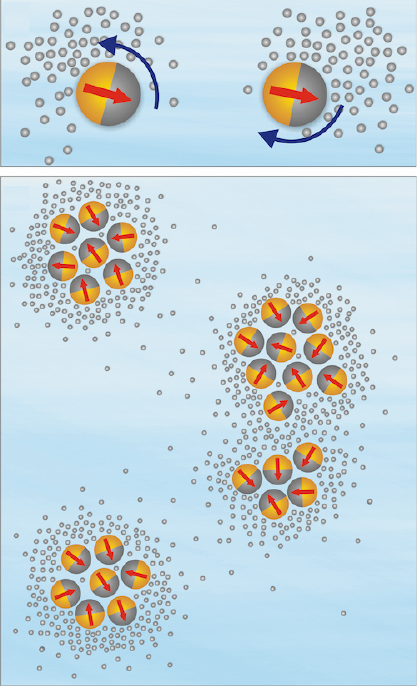}
\label{consfig} 
\end{figure}

In this Account, we review recent progress in the theoretical description of synthetic chemotaxis mainly based on simulations and field theoretical descriptions.  
We will begin with single motile particles leaving chemical trails behind with which they interact themselves, 
leading to effects like self-trapping or self-avoidance. 
Besides these self-interactions, in ensembles of synthetic motile particles each particle also responds to the chemicals produced by other 
particles, inducing chemical (or phoretic) cross-interactions. 
When these interactions are attractive, they commonly lead to clusters, even at low particle density. These clusters may either proceed towards 
macrophase separation, resembling Dictyostelium aggregation,
or, as shown very recently, lead to dynamic clusters of 
self-limited size (dynamic clustering) as seen in experiments in autophoretic Janus colloids.

Besides the classical case where chemical interactions are attractive, 
this Account discusses, as its main focus, repulsive chemical interactions, which can create a new and less known avenue to pattern formation in active systems 
leading to a variety of pattern, including clusters which are surrounded by shells of chemicals, 
travelling waves and more complex continously reshaping patterns. 
In all these cases ``synthetic signalling'' can crucially determine the collective behavior of synthetic microswimmer ensembles 
and can be used as a design principle to create patterns in motile active particles. 


\section{\rule{10pt}{10pt} Introduction}
To find food and to avoid toxins, microorganisms are equipped with a remarkable navigation machinery, which allows them 
to
sense (or ``smell'') certain chemicals to which they respond by moving either up the chemical 
gradient (chemoattraction) or down it (chemorepulsion or negative chemotaxis). 
Besides chemotaxis, microorganisms show taxis also to many other gradients, such as temperature (thermotaxis) \cite{Mori1995}, light intensity (phototaxis) \cite{Witman1993,Berry2000}
or viscosity (viscotaxis) \cite{Daniels1980,Liebchen2018}.

Remarkably, many microorganisms can produce the chemicals to which they respond themselves and use 
chemotaxis to communicate with each other. 
This communication, called signalling, allows microorganisms (and cells) to 
coordinate their motion and gene expression, 
which is crucial for a large variety of biological processes: 
for example, it 
allows the sperm to find the egg, thus preceding mammalian life \cite{Murray2009,Eisenbach2004}. 

\begin{figure}
\includegraphics[width=0.95\textwidth]{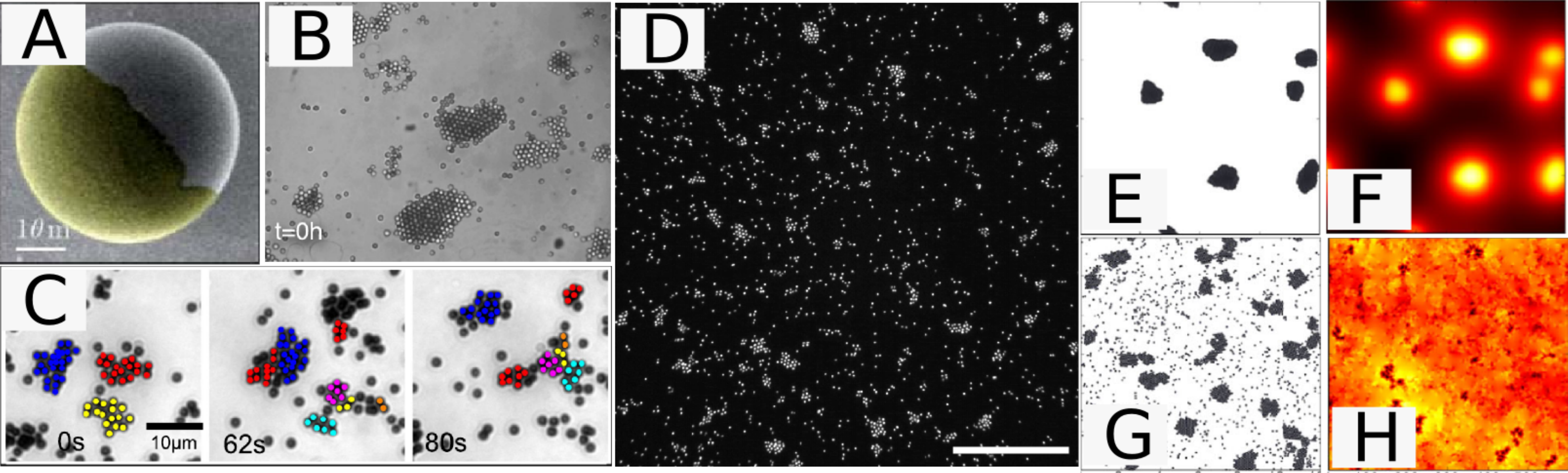}
\caption{Panel A shows an active Janus colloid which catalyzes a chemical reaction on part of its surface and swims by a 
phoretic mechanism in the self-produced chemical gradient (Reproduced with permission from ref. \cite{Volpe2011}. Copyright 2011 Royal Society of Chemistry).
The same gradients also act on other colloids, leading to chemical cross-interactions probably underlying 
dynamic clusters commonly seen in low density ensembles of Janus colloids
Figs. B,C (Reproduced with permission from ref.\cite{Buttinoni2013} Copyright 2013 American Physical Society and from ref.~\cite{Theurkauff2012}. Copyright 2012 
American Physical Society), Fig. D from ref.\cite{Ginot2018}. 
Panels E-H (Reproduced with permission from ref.~\cite{Liebchen2017phoretic}. Copyright 2018 American Physical Society)
show clusters seen in simulations of colloids with attractive (E,F) and (partly) repulsive (G,H) 
chemical cross-interactions \cite{Liebchen2017phoretic}, discussed later in the text.}
\label{dynclusters_exp} 
\end{figure}

An example illustrating collective behavior based on signalling is provided
by Dictyostelium cells; when starving, they produce certain chemicals (cAMP)
whose ``smell'' attracts other cells inducing a positive feedback loop: 
a local cell accumulation yields an enhanced chemical production, 
further enhancing the initial smell attracting other cells etc. 
Overall, this results in a collapse of Dictyostelium 
cells followed by the formation of 
a multicellular object (sporulation) allowing the cells to survive long starvation periods \cite{Bonner1947,Gerisch1968,Eisenbach2004}; see \cite{Eidi2017} for a recent model 
on signalling Dictyostelium.
(The same aggregation mechanism is now also established for {\it E.coli} showing positive chemotaxis to self-produced autoinducers \cite{Laganenka2016}).
This mechanism is captured in 
the classical Keller-Segel model \cite{Keller1970,Keller1971a} as we will quantify below. 

\begin{figure}
\includegraphics[width=0.3\textwidth]{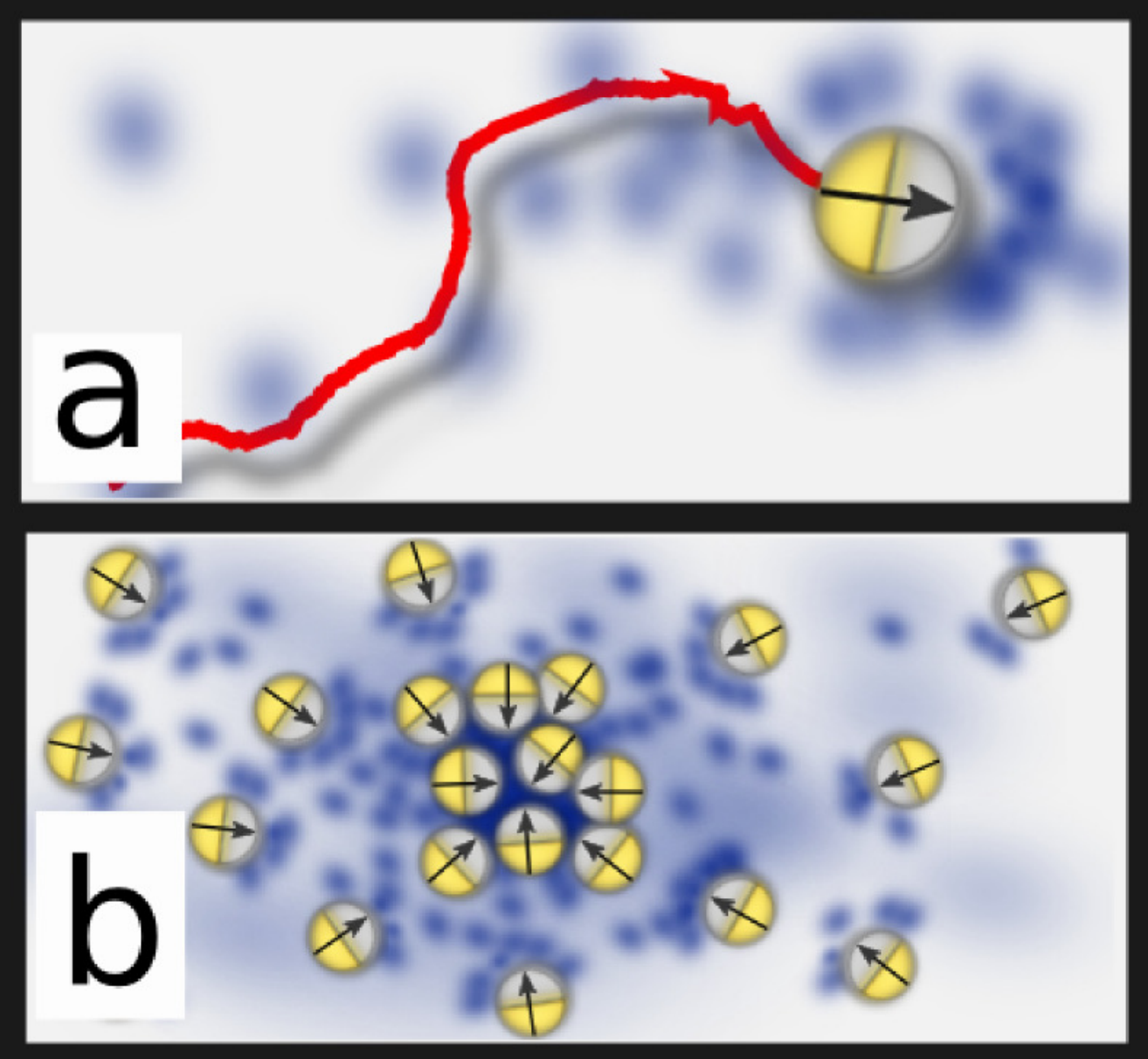}
\caption{a.) Cartoon of a self-propelled Janus colloid producing certain chemicals causing 
chemical self-interactions. b.) Cartoon of the Keller-Segel instability for Janus colloids responding both 
to self-produced chemicals and chemicals produced by other Janus colloids (chemical cross-interactions): 
a local accumulation of colloids leads to a locally enhanced chemical production which in turn biases the motion of further active particles to move up the chemical gradient, 
supporting the 
original colloidal accumulation and so on. Fig. from\cite{Liebchen2016}.} 
\label{KSmech}
\end{figure}

Recent advances in synthetic microswimmers \cite{Marchetti2013,Bechinger2016} have revealed a remarkably close
synthetic analogon to biological chemotaxis:
Phoretic Janus-colloids which are half-coated with a 
catalytic material like gold or platinum catalyze chemical reactions in the solute surrounding them (Figs.~\ref{dynclusters_exp}A 
,\ref{KSmech}a). 
This results in a chemical gradient 
across the colloids' surfaces, driving them forward by 
diffusiophoresis or a similar mechanism based on interactions of the solutes and the interfacial layers of the colloids. 
Interestingly, the self-produced chemical gradients do not only lead to self-propulsion, but also act on other Janus colloids 
(Janus colloids can of course not distinguish self-produced
chemicals from those produced by other colloids); Fig.~\ref{KSmech}b.
\\The analogy of chemically interacting Janus colloids and signalling microorganisms holds true even formally: for example, the 
same Keller-Segel equations which describe the aggregation of microorganisms \cite{Tindall2008} apply to 
(chemically interacting) Janus colloids
\cite{Meyer2014,Saha2014,Pohl2015,Liebchen2015} opening the perspective to 
explore signalling-induced pattern formation as observed in biological systems in 
a minimal synthetic environment.
Note that synthetic signalling is not restricted to 
chemical interactions but 
can occur e.g. also in self-thermophoretic and self-electrophoretic colloids where particles interact via other self-produced
phoretic fields (e.g. temperature, electric field) 
which they use for swimming
\cite{Jiang2010,Kroy2016,Bickel2014,Geiseler2017,Io2017}. We thus use the terms phoretic interactions and synthetic signalling interchangeably in the following.
 
Here, the ``interaction fields'' are materialistic and follow the motion of the Janus colloids non-instantaneously 
inducing memory or delay effects allowing particles to interact with their own past (they leave chemical trails behind) and with the history of other particles. 
Such delay effects should also be relevant in self-propelled oil droplets showing chemotaxis 
with respect to 
micellar surfactant gradients (chemoattractive response) and to empty micelles which they leave in their wake \cite{Maass2016,Jin2017} and 
camphor boats \cite{Nakata2005,Kohira2001} leaving slowly decaying chemical trails in their wakes.

While we are still at the beginning of understanding the plethora of physical phenomena made possible by interactions based on synthetic chemotaxis, 
it is likely that they play a key role e.g. for the typical collective behavior of autophoretic microswimmers. 
In particular,  
dilute suspensions of Janus colloids of only 3-10 per cent packing fraction, spontaneously form so-called living clusters in experiments 
(also called dynamic clusters)
\cite{Theurkauff2012,Palacci2013,Buttinoni2013,Ginot2018}.
Remarkably, as a hallmark of their nonequilibrium nature, these clusters are intrinsically dynamic and continuously break up and reform.

In typical active systems, phoretic interactions compete with steric short range repulsions 
and hydrodynamic interactions, making it important to understand which interactions dominate for a given active setup \cite{Liebchen2018PIs}: 
from the exploration of minimal active matter models, it is known that 
the combination of steric short range repulsions and motility alone can lead to 
spontaneous aggregation of particle ensembles \cite{Cates2015} (``motility induced phase separation'')). However, 
this requires packing fractions $\gtrsim 30\%$ to occur sponaneously from a uniform phase \cite{Stenhammar2013} or a large nucleation seed \cite{Levis2017}. 
Similarly, hydrodynamic interactions 
are known to often play an important role not only for microorganisms \cite{Saintillan2008} but also for the collective behavior of colloids
at moderate to high density
\cite{Nagele1996,Zottl2014}.
Conversely, at low density where many experiments with Janus colloids are performed, chemical (phoretic) interactions are probably more important in many 
setups
\cite{Palacci2013,Singh2017,Liebchen2018PIs}. In particular, these interactions generically 
destabilize the uniform phase in Janus colloids (and dimers) 
even at low area fraction \cite{Liebchen2017phoretic,Liebchen2018PIs,Colberg2017,Robertson2018}
and can lead to clusters of self-limited size \cite{Pohl2014,Liebchen2015,Liebchen2017,Liebchen2018PIs} as in experiments 
\cite{Theurkauff2012,Palacci2013,Buttinoni2013,Ginot2018}. 
On the experimental side, long-range attractive phoretic interactions have been measured both in Janus colloids forming dynamic clusters \cite{Palacci2013}
and microgears \cite{Aubret2018}
and in low density active-passive mixtures \cite{Singh2017}, where passive particles cluster 
around active seeds showing various (local) crystal structures (hexagonal, square lattice, glassy configurations).
Finally, experiments in ${\rm Ag}_3{\rm PO}_4$-microparticles which lead to schooling patterns and allow for transitions to patterns involving large exclusion zones 
seem to be also dominated by phoretic interactions \cite{Ibele2009,Duan2013}.
\\\textbf{Organization of the article:} 
In the following, we discuss recent progress regarding the modelling of chemical interactions in synthetic microswimmers. Here, we briefly discuss 
chemotaxis of individual particles which can self-produce the chemicals to which they respond (autochemotaxis) and will then consider 
their collective behavior, 
which can lead to spontaneous aggregation (Keller-Segel model).
We will then discuss
chemotaxis in active particles, the key topic of this article. 
Here, chemical gradients create both an effective force and an effective 
torque acting on the center of mass and the swimming direction of the active particles. With the {\it Phoretic Brownian Particle} (PBP) model, 
we will 
discuss a minimal model describing the collective dynamics of chemically interacting active particles, 
leading to patterns, including 
spontaneous aggregation, dynamic clustering, travelling waves and spiral patterns.
Finally, we will briefly discuss the physics of chemically interacting chiral active particles and beyond the PBP model.



\section{\rule{10pt}{10pt} Chemotaxis in Passive Particles}
\subsection{Modelling Single Particle Chemotaxis}
Let us consider a simple model for chemotaxis, describing the dynamics of an isotropic 
overdamped Brownian particle with center of mass coordinate ${\bf r}_1$ which couples to the gradient of a field 
$c({\bf r},t)$. The corresponding Langevin equation reads
\1 
 \dot {\bf r}_1 (t)  = \beta_D \nabla c({\bf r}_1(t),t) + \sqrt{2D}{\bm \xi}(t) \label{eom1} 
\2
Here, we call $c$ the 
``chemical field'' for concreteness but keep in mind that $c$ may also represent e.g. a temperature or light intensity field, depending on the 
type of taxis we consider. 
$\beta_D$ is the (chemo)tactic coupling coefficient, ${\bm \xi}(t)$ represents Gaussian white noise of zero mean and unit variance and $D$ is the diffusion coefficient of the particle. 
If $\beta_D>0$, the particle moves towards high chemical concentration and shows {\it chemoattraction} (or positive chemotaxis); if $\beta_D<0$, the particle moves down the 
chemical gradient representing {\it chemorepulsion} (or negative chemotaxis).

As we have discussed above, many microorganisms self-produce the chemical to which they respond, say with a rate $k_0$. 
This is called (positive or negative) {\it autochemotaxis}.
We describe the corresponding chemical dynamics by a diffusive evolution equation with a source representing chemical production by the particle
\1 \label{chem}
\dot c({\bf r},t) = D_c \Delta c({\bf  r}, t) + k_0\delta ( {\bf r} -{\bf r}_1)- k_d c({\bf r}, t)
\2
where $D_c$ is the chemical diffusion coefficient. 
The sink term, led by the rate coefficient $k_d$ represents chemical evaporation, occurring e.g. due to secondary chemical reactions taking place in the 
underlying solvent. 
For autochemoattraction ($\beta_D>0$) where particle produces a chemical to which it is attracted, Eqs.~(\ref{eom1},\ref{chem}) can lead to self-trapping in one and two dimensions 
(but not in three) \cite{Tsori2004}
which can be permanent ($k_d=0$ \cite{Tsori2004}) or transient ($k_d>0$ \cite{Grima2005,Grima2006}) and is opposed by noise \cite{Sengupta2009}.
Conversely, autochemorepulsion ($\beta_D<0$) can lead to self-avoidance \cite{Grima2005,Grima2006}.
A system of two chemotactic particles, A and B, where A is attracted by the chemical released by B, and B is repelled by the chemical produced by A has been 
explored in \cite{Sengupta2011} and forms a close analogue to a predator-prey system featuring nonreciprocal interactions. 
For more details on single and two-particle chemotaxis we refer the reader to\cite{Liebchen2018bookchapter}. 
Here, we mainly focus on collective behavior.

\subsection{Collective Behaviour -- Keller-Segel Model}
We now consider ensembles of diffusive particles interacting via self-produced chemical fields (signalling). 
Here, the dynamics of each particle couples to the chemicals 
produced by \emph{all} $N$ particles in the system. Formally, to describe collective behavior, we can replace 
${\bf r}_1(t)\rightarrow {\bf r}_i(t)$ ($i=1,..,N$) in (\ref{eom1}) and 
of $k_0\delta\left({\bf r}-{\bf r}_1(t) \right) \rightarrow \Sum{i=1}{N} k_0\delta\left({\bf r}-{\bf r}_i(t) \right)$ in (\ref{chem}).
It is convenient to describe collective behavior using a particle density field 
$\rho({\bf r},t)=\Sum{i=1}{N} \delta\left({\bf r}-{\bf r}_i(t)\right)$ coupled to the chemical density. 
The exact Smoluchowski-equation for $\rho$ coupled to an evolution 
equation for $c$ reads: 
\ea
\dot \rho &=&-\nabla \cdot (\beta_D \rho \nabla c) + D \nabla^2 \rho \label{Ks1}\\
\dot c &=& D_c \nabla^2 c + k_0 \rho - k_d c \label{Ks2}
\ee
These equations represent the classical Keller-Segel model (see \cite{Tindall2008} for variants of this model).
One obvious solution is $\left(\rho,c\right)=\left({\rho_0,k_0\rho_0/k_d}\right)$ representing a uniform disordered phase. 
Performing a linear stability analysis of this phase predicts a criterion for the onset of structure formation, which reads:
\1 
k_0 \rho_0 \beta_D > D k_d \label{KSinst}
\2 
This is the Keller-Segel instability which occurs for chemoattraction ($\beta_D>0$)
and is based on a positive feedback between 
particle aggregation and chemical production.
Following the instability criterion (\ref{KSinst}) and strong overall chemical production ($\sim k_0 \rho_0$), 
strong response to the chemical ($\sim \beta$) support the emergence of an instability, whereas fast evaporation (decay) 
and fast particle diffusion oppose it. 
The Keller-Segel instability, typically leads to clusters (Fig.~\ref{dynclusters_exp} E) and colocated chemical clusters (F)
which both coarsen (and coalesce) and lead to one dense 
macrocluster at late times, resembling a gravitational collapse \cite{Tsori2004,Sire2004,Bleibel2011}. 
For weak chemotactic coupling
such a collapse can be prevented by reproduction processes which are naturally present in microorganisms, but not for strong coupling \cite{Gelimson2015}.

%

\section{\rule{10pt}{10pt} Chemotaxis in Active Particles}
\subsection{Active Particles in an Imposed Chemical Gradient}
Many chemotactic particles, from cells and microorganisms to synthetic Janus colloids self-propel, i.e. they move autonomousely.
When exposed to a chemical gradient, these anisotropic particles displace (change of velocity) 
and align with (or against) the chemical gradient (change of self-propulsion direction). 
Effectively, chemical gradients cause both a force and a torque on active particles. 
It is convenient 
to describe the motion of an overdamped chemotactic particle self-propelling with a velocity $v_0$ (independently of chemotaxis) 
in a direction ${\bf p}=(\cos\theta,\sin\theta)$ in two dimensions
by the following evolution equations
\ea
 \dot{\bf r} (t)  &=& v_0 {\bf p} + \beta_D \nabla c({\bf r}(t), t) + \sqrt{2D}{\bm \xi}(t) \label{reom}\\
 \dot\theta (t)  &=& \beta {\bf p} \times \nabla c({\bf r}(t), t) + \sqrt{2D_r}{\eta}(t) \label{theom} 
\ee
where ${\bf a} \times {\bf b} = a_1 b_2 - a_2 b_1$ represents the 2D-cross-product.
Here, the chemotactic drift, $\beta_D \nabla c$, changes the speed of the particle in response to the imposed chemical gradient and 
the alignment term, $\beta \nabla c$, changes the particle orientation (swimming direction): 
when $\beta>0$ the particle (i.e. its swimming direction) turns up the chemical gradient
(chemoattraction) whereas for $\beta<0$ the particle turns down the gradient (chemorepulsion); see the left two columns in Fig.~\ref{class}.
This alignment occurs in competition with rotational Brownian diffusion 
due to 
collisions with solvent molecules, described by the rotational diffusion coefficient $D_r$ and Gaussian white noise $\eta$ with zero mean and unit variance.
\\Synthetic chemotaxis of active particles has been first observed with microrods in externally imposed chemical gradients
in ref.~\cite{Hong2007}. Later synthetic taxis has
discussed also for thermophoretic Janus colloids in externally imposed temperature gradients \cite{Bickel2014,Cohen2014}.
In \cite{Bickel2014} the temperature gradient induces an effective polarization of the particles, showing that the temperature gradients (also) act on the 
orientational degrees of freedom of the particle; and 
in \cite{Cohen2014}, particles shield each other from the light (heat) source, 
leading to the formation of a moving swarm. Recently,
synthetic phototaxis has also been observed in artificial microswimmers which has beed used to create a directed particle transport in a ratchet-shaped light intensity field \cite{Lozano2016}.
\\An active version of {\it autochemotactic} particles interacting with their own trails has been discussed in 
~\cite{Taktikos2011,Kranz2016} and typical trajectories of two (and more) of these walkers have been discussed in \cite{Taktikos2012}.

\subsection{Collective Behaviour: the Phoretic Brownian Particle Model}
The chemical field produced by an autophoretic colloid decays slowly with increasing distance, as $1/r$ (gradients as $1/r^2$), 
if $k_d$ is not too large, and may thus create a significant effective torque acting on the orientation of all other particles in the system.  
Ensembles of autophoretic swimmers therefore represent a system where each particle swims in self-produced gradients 
in a direction which is influenced by the chemical gradients produced by all other particles in the system. 
This leads to a complex collective behavior which can be described, in a minimal form, by the
{\it Phoretic Brownian Particle} (PBP) model \cite{Liebchen2017phoretic}.
This model is based on equations of motion for $N$ chemotactic swimmers, each described by, 
Eqs.~(\ref{reom},\ref{theom}), coupled to each other via a self-produced chemical field which evolves following the 
chemical diffusion equation.
To define the PBP equations, we consider the following simplifications of Eqs.~(\ref{reom},\ref{theom}).
\\(i) As commonly assumed in active matter passive diffusion plays only a minor role, since the 
combination of rotational diffusion and self-propulsion effectively leads to a much stronger effective diffusion
$\sim v_0^2/D_r\gg D$; i.e. we set $D\rightarrow 0$ in (\ref{reom}). 
\\(ii) For simplicity, the PBP model also neglects the chemotactic drift ($\beta_D\rightarrow 0$),
allowing us to capture most aspects of chemically interacting microswimmers in a simplified way 
(for cases when drift and alignment are both attractive or both 
repulsive).
Generally, drift effects may also be important (or even dominant), e.g. when the chemical interactions are attractive \cite{Liebchen2018PIs}. 
\\We can now define the PBP model for $N$ self-propelled particles moving with identical 
self-propulsion velocities $v$ in directions ${\bf p}_i=\left(\cos \theta_i(t),\sin\theta_i(t)\right)$ $i=1,..,N$, 
which change, due to rotational diffusion and chemical torques. 
\ea
\dot {\bf r}_i &=& v_0 {\bf p}_i \\
\dot \theta_i &=& \beta {\bf p}_i \times \nabla c({\bf r}_i) + \sqrt{2D_r} \xi_i (t);\quad i=1..N
\ee
To account for possible anisotropic chemical production, occurring e.g. for Janus colloids, we slightly generalize the evolution equation for the chemical:
\1 \label{chemaniso}
\dot c({\bf r},t)= D_c \nabla^2 c({\bf  r}, t) - k_d c({\bf r}, t) + \Sum{i=1}{N}\oint {\rm d} {\bf x}_i\delta ( {\bf r} -{\bf r}_i(t) - R_0 {\bf x}_i)\sigma({\bf x}_i)
\2
The integral is over the 3D surface of the (spherical) particles with radius $R_0$ and $\sigma({\bf x}_i)$ is the (nonuniform) production rate density on the particle surface.
Specifically for Janus colloids, we have $\sigma({\bf x}_i)=k_0/(2\pi R_0^2)$ on the catalytic hemisphere and zero elsewhere.
Since we are mainly interested in understanding the onset of pattern formation at low densities we neglect steric short range repulsions among the particles
for simplicity (but will include them in particle based simulations).

\textbf{Limit of fast chemical dynamics and the Active Attractive Alignment Model:} 
Seemingly, when assuming that the chemical diffuses much faster than the particles ($D_c \gg v_0^2/D_r$) 
we may treat the chemical field as a fast variable, enslaved by the motion of the particles, i.e. to set $\dot c \rightarrow 0$ in Eq.~(\ref{chem}).
This approach leads to the Active Attractive Alignment model (when accounting also for drift effects), 
which has been developped very recently in \cite{Liebchen2018PIs} and focuses on attractive 
phoretic interactions 
The AAA model naturally leads to dynamic clustering at low densities in accordance with experiments. 
In general however, the limit of fast chemical diffusivity is dangerous for Janus colloids, 
since the chemical coupling coefficients $\beta$ (and $\beta_D$) turn out to be proportional to $D_c$ such that 
$c$ cannot be treated as a fast variable \cite{Liebchen2017phoretic}.
Thus, delay or memory effects may be important even for large $D_c$, which turns out to be particularly important for 
repulsive phoretic interactions as we will discuss below in detail. 

\subsection{Field Theory of the Phoretic Brownian Particle Model}
\begin{figure}
\includegraphics[width=0.75\textwidth]{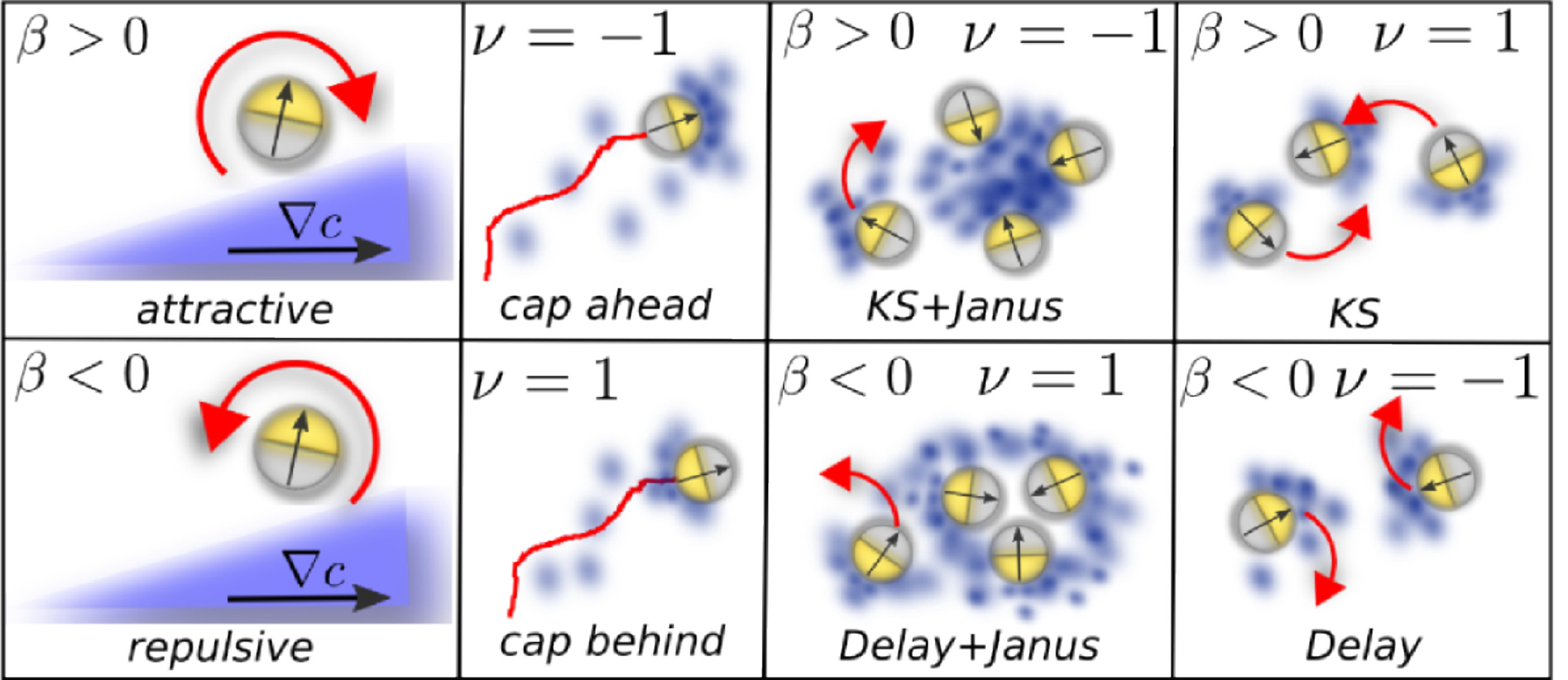}
\caption{Classification of Janus colloids and the instabilities to which they can lead in the PBP model. Here, KS represents the Keller-Segel 
instability and Janus and Delay stand for the Janus instability and the delay induced instability discussed further below in the text. 
Reproduced with permission from ref.~\cite{Liebchen2017phoretic}. Copyright 2017 American Physical Society.}
\label{class} 
\end{figure}

To understand the collective behavior of phoretically interacting active particles, it is useful to coarse grain the PBP model. 
Since active particles are nonisotropic, but have a direction of motion, a field theory for their collective behavior naturally 
involves not only the particle density field $\rho({\bf r},t)=\Sum{i=1}{N} \delta({\bf r}-{\bf r}_i(t))$ but also a polarization density field ${\bf w}({\bf r},t)$
 defined as ${\bf w}=\Sum{i=1}{N}{\bf p}_i \delta({\bf r}-{\bf r}_i(t))$. Its magnitude is a measure for the number of aligned particles around position ${\bf r}$
and its direction describes the average self-propulsion direction.
Following \cite{Liebchen2017}, for moderate 
deviations from isotropy we obtain
\ea
\dot \rho &=& - {\rm Pe} \nabla \cdot {\bf w} 
\label{PBPf1} \\
\dot {\bf w} &=& -{\bf w} + \4{B\rho}{2}\nabla c - \4{{\rm Pe}}{2}\nabla \rho  + \4{{\rm Pe}^2}{16}\nabla^2 {\bf w} - \4{B^2 |\nabla c|^2}{8} {\bf w} \nonumber \\
&+&
\4{{\rm Pe} B}{16} \left(3 (\nabla {\bf w})^{\rm T}\cdot \nabla c - (\nabla c \cdot \nabla){\bf w} - 3 (\nabla \cdot {\bf w})\nabla c \right)
\label{PBPf2} \\ 
\dot c &=& \mathcal{D} \nabla^2 c  + K_0 \rho + \nu \4{K_0}{2}\nabla \cdot {\bf w} - K_d c \label{PBPf3}. 
\ee
Here, we have introduced time and space units as $t_u=1/D_r$ and $x_0=R_0$, 
leading to the following six dimensionless control parameters
(i) the Peclet number ${\rm Pe}=v_0/(R_0 D_r)$; (ii) $B=\beta/(D_r R_0^4)$; 
(iii), (iv) $K_0=k_0/D_r; K_d=k_d/D_r$, (v) $\mathcal{D}=D_c/(R_0^2 D_r)$ and (iv) the dimensionless quasi-two dimensional density $\rho_0=x_u^2 N/L^2$.
Finally, $\nu$ determines the anisotropy in the chemical production; we have $\nu=0$ for isotropic chemical production 
and specifically for Janus colloids, we have $\nu=1$ for those moving with the catalytic cap in the back and $\nu=-1$ for 
Janus swimmers moving cap-ahead; compare Fig.~\ref{class}.

\subsection{Active Keller-Segel Model}
One interesting limiting case of Eqs.~(\ref{PBPf1}--\ref{PBPf3}) is obtained by assuming that particles reorient quasi-instantaneously ($\dot {\bf w}\rightarrow {\bf 0}$)
and by neglecting nonlinear terms which will be important only far from the uniform phase as well as second order gradients 
in Eq.~(\ref{PBPf2}), i.e. ${\bf w}\approx B\rho \nabla c/2-{\rm Pe}\nabla \rho/2$. Plugging this expression into Eq.~(\ref{PBPf1}) and neglecting the 
anisotropy in the production ($\nu=0$) yields the 
Keller-Segel equations (\ref{Ks1},\ref{Ks2}) now applying to active particles and written in dimensionless units.
Thus, the apparent similarity between signalling microorganisms and synthetic microswimmers holds true also formally. 
The formal analogy allows us to immediately write down the instability criterion as
\1 \4{K_0 \rho_0 B}{K_d {\rm Pe}}>1. \label{actKSinst}\2
It can be shown\cite{Liebchen2017phoretic} that despite the approximations which we have used to derive the active Keller-Segel model, this instability 
criterion holds true exactly for Eqs.~(\ref{PBPf1})--(\ref{PBPf3}).
The Keller-Segel instability leads to clusters of active particles and colocated clusters of the self-produced chemical which grow in the coarse of the 
time due to coarsening and cluster-coalescence (see Fig.~\ref{dynclusters_exp} E,F).
In physical units the instability criterion translates to $k_0 \rho_0 \beta/(k_d v_0)>1$ showing that strong production and alignment up chemical gradients favors instability of the 
uniform phase. However, it also suggests that self-propulsion opposes the instability 
which seems paradoxical, since in the absence of self-propulsion, particles in the PBP model do not move at all and hence, should not cluster. 
We will resolve this paradoxon in the next paragraph. 

\subsection{Parameter Collapse and Universality}
As discussed in the introduction, phoretic microswimmers move by catalyzing certain chemical reactions in a bath on part of their surface only, which drives them forward. 
The same gradients also act on other phoretic swimmers and bias their swimming direction. 
It has been shown that this situation allows us to express the (dimensionless) phoretic cross coupling coefficient $B$,
through parameters determining the single-particle swimming speed \cite{Liebchen2017phoretic} as 
\1 B \approx \4{4\pi s {\rm Pe} \mathcal{D}}{K_0} \label{Blink} \2
where $s=1$ represents chemoattraction and $s=-1$ represent chemorepulsion. 
This expression for $B$ can now be used to strongly simplify the Keller-Segel instability for active particles. 
Combining (\ref{Blink}) with (\ref{actKSinst}) yields the following instability criterion $6\mathcal{D}\rho_0/K_d>1$ or with the quasi-2D area 
fraction $f=\pi \rho_0$ ($f=\pi R_0^2 \rho_0$ in physical units):
\1
6\mathcal{D}f>K_d \label{collKS}
\2
In physical units, criterion (\ref{collKS}) can be written as $D_c/(R_0^2 k_d)>1/(6f)$
where $D_c/(k_d R_0^2)\sim 1-10$ \cite{Liebchen2018PIs} suggests that the Janus instability 
might occur at low area fractions of a few per cent only. In real Janus colloids with attractive phoretic interactions, 
drift effects further support this instability. 

\subsection{Chemorepulsive Route to Pattern Formation}
\begin{figure}
\includegraphics[width=0.4\textwidth]{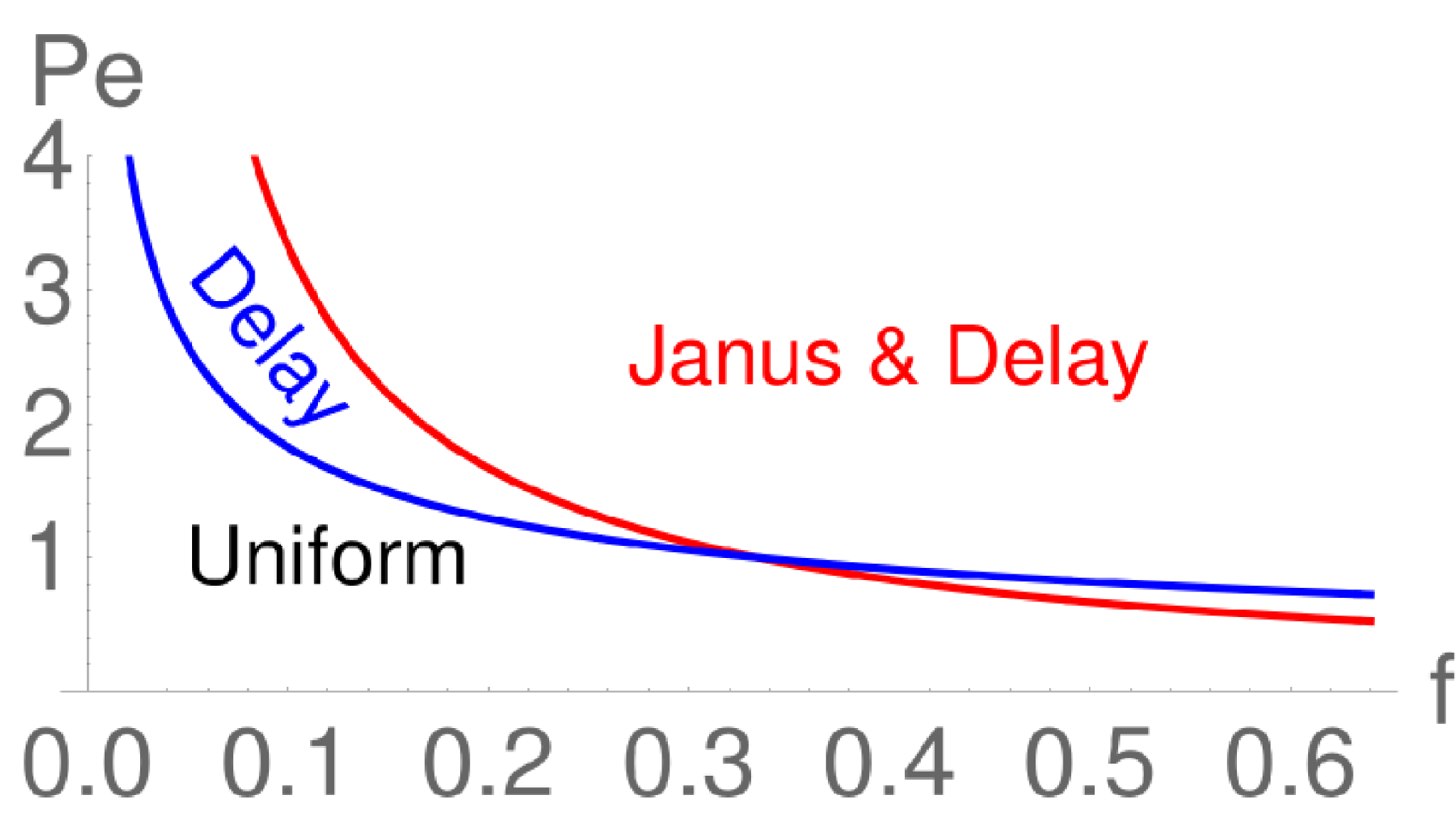}
\caption{Phase diagram of the Phoretic Brownian Particle model for $\beta<0$ (repulsive phoretic interactions), $k_d=0$ suggesting that
not only attractive phoretic interactions but also repulsive ones destabilize the uniform phase far below the the onset of motility-induced phase separation 
occurring e.g. in the Active Brownian particle model.
Reproduced with permission from ref.~\cite{Liebchen2017phoretic}. Copyright 2017 American Physical Society; see ref.\cite{Liebchen2017phoretic} for further details.}
\label{instdiag} 
\end{figure}

Remarkably, not only attractive signalling can destabilize the disordered uniform phase, but also 
repulsive signalling, which in fact, creates
a proper route to pattern formation leading to a large variety of possible structures including clusters and wave patterns. 
A general linear stability analysis of Eqs.~(\ref{PBPf1})--(\ref{PBPf3}) 
for $K_d\ll 1$ leads, together with 
Eq.~(\ref{Blink}) to the following instability criterion of the uniform phase for $B<0$
\1
3{\rm Pe} f \left(\4{\nu}{2} + {\rm Pe}\right)\gtrsim 1 \label{repinstab}
\2
Note that (\ref{repinstab}) represents a massive collapse of the parameter space of the PBP model. 
From originally six dimensionless paramters, it depends only on two parameters ${\rm Pe}$ and $f$ (and $\nu$ which is of order 1) allowing to compare the 
phase diagram of the PBP model (Fig.~\ref{instdiag}) with that of the widely used active Brownian particle model. 
Since we have ${\rm Pe}\sim 20-200$ 
for typical autophoretic Janus swimmers \cite{Theurkauff2012,Palacci2013,Buttinoni2013},
this criterion shows that not only attractive chemical interactions but also repulsive ones generically destabilize the uniform phase in autophoretic microswimmers. 
Thus, the very fact that Janus colloids swim obliges them to form patterns -- 
even at very low density \cite{Liebchen2017phoretic}, where the Active Brownian particle model predicts stability of the uniform phase only, but experiments 
show dynamic clustering \cite{Theurkauff2012,Palacci2013, Buttinoni2013,Ginot2018}. 
(This remains true in the presence of chemical evaporation ($K_d\neq 0$), 
up to $K_d \sim {\rm Pe}^2 f$.)

\begin{figure}
\includegraphics[width=0.75\textwidth]{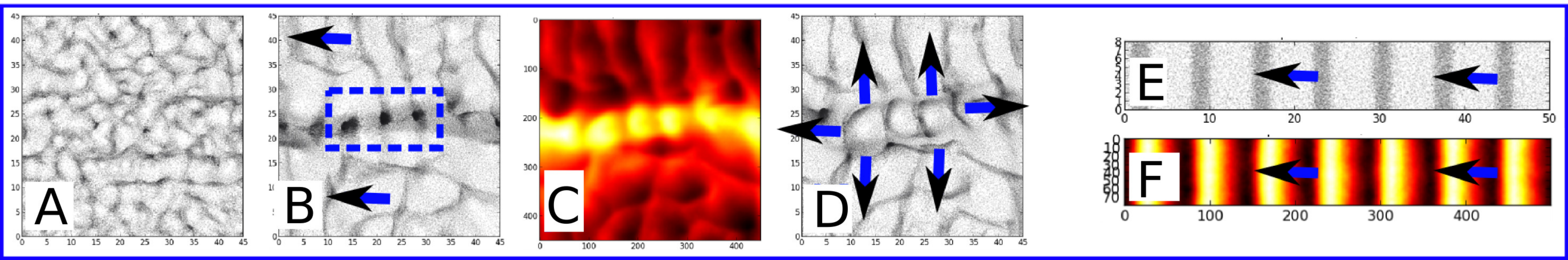}
\caption{Simulations of the Phoretic Brownian Particle model for repulsive phoretic interactions 
including additional short-ranged steric repulsions 
among the particles.
The figure shows colloidal waves
(A) pursued by self-produced phoretic waves caging the colloids in dense clusters (B); these clusters act as enhanced phoretic producers
leading to phoretic clusters (C) which drive colloids away, and induce escape waves (D). At late times, these wave patterns may settle into
regular moving bands of colloids closely followed by phoretic waves (E,F).
Reproduced with permission from ref. \cite{Liebchen2017phoretic}. Copyright 2017 American Physical Society.} 
\label{instabmech} 
\end{figure}

The instability criterion Eq.~(\ref{repinstab}) is a sum of two terms representing independent physical mechanisms, 
illustrated in Fig.~\ref{instabmech} and discussed in the following (for more details see \cite{Liebchen2015}).
\begin{itemize} 
\item \textbf{Janus instability:} The first part of the criterion reads
$3{\rm Pe}f\nu>2$. Since $\nu=0$ for isotropic chemical production on the surface of the particles, this criterion can only be 
fulfilled for anisotropic chemical production; hence it is called, the 
(``\emph{Janus instability}'').
The Janus instability leads to clusters of finite size. 
The underlying physical mechanism is illustrated in Fig.~\ref{instabmech} A.
(Note that anisotropic chemical production effectively leads here to anisotropic interactions among Janus particles; 
a model involving direct anisotropic interactions has been considered in \cite{Pu2017} and leads to similar clusters.)
\item \textbf{Delay-induced instability:}
The second criterion $3{\rm Pe}^2 f>1$ in Eq.~(\ref{repinstab}) represents an 
oscillatory instability and creates wave patterns. 
This instability is based on a delay in the response of the colloids to changes 
in the chemical field an is therefore referred to as the ``\emph{delay-induced instability}'', which is 
detailed in Fig.~\ref{instabmech} B. 
\end{itemize}

\begin{figure}
\includegraphics[width=0.99\textwidth]{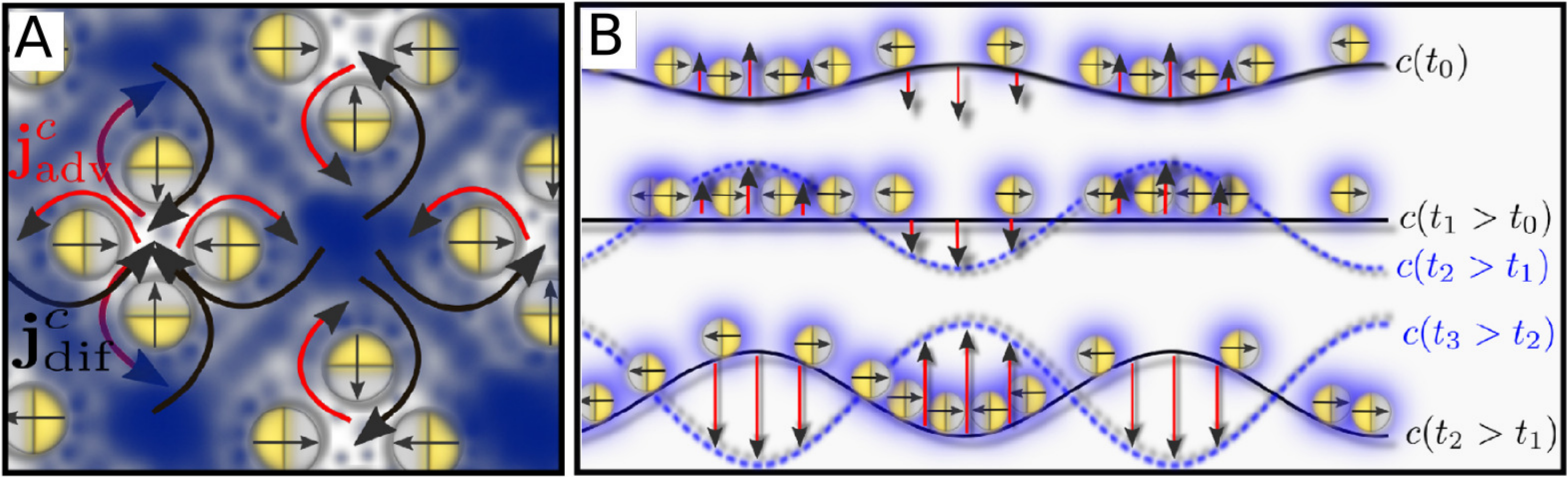}
\caption{Instability mechanisms in repulsively signalling active particles:
A.) Janus instability based on anisotropic chemical production of colloids on their 
surface leading to clusters of self-limited size which are 
surrounded by self-produced shells of high chemical density which keeps the 
particles within the cluster together and keeps other colloids away.
B.) Delay induced instability. Upper panel: Chemorepulsive colloids accumulate in the 
minima of a weak initial fluctuation of the uniform chemical field. Middle panel: Chemical 
production by the colloids opposes the original fluctuation, but does not stop at 
uniform chemical density but overshoots, due to a finite response time of the 
colloids. This reverses the original profile and may amplify (lower panel) triggering 
an oscillatory instability.
Reproduced with permission from ref. \cite{Liebchen2015}. Copyright 2015 American Physical Society.}
\label{reppat}
\end{figure}

The Janus and the delay-induced instability create a rich panorama of patterns. 
In most cases the delay-induced instability dominates over the Janus instability and leads to continuously evolving patterns\cite{Liebchen2015,Liebchen2017phoretic}.
A typical example is shown in Fig.~\ref{reppat}. It involves colloidal waves
pursued by self-produced phoretic waves. 
When these waves collide frontally, the pursuing phoretic waves act as cages for the chemorepulsive particles 
and morph them into a cluster (Fig.~\ref{reppat} B).
The high particle density
within such a cluster induces an enhanced chemical production resulting in colocated phoretic clusters (C) which in turn expel the chemorepulsive colloids.
This expulsion may occur rather suddenly and may initiate 
new colloidal ring
waves leaving low density regions at the locations of the
former clusters (D). These waves continue colliding, create new clusters and so on. 
At very late times, these waves may settle down into a regular pattern of travelling waves (E) pursued by
co-travelling chemical waves (F).
When two waves colloids rather than frontally, they often create moving clusters as discussed in some detail in \cite{Liebchen2017phoretic}.
Recent experiments in ensembles of thermophoretic active colloids \cite{Io2017} 
find corresponding travelling clusters, which might have a similar origin.

While these patterns mainly hinge on the delay-induced instability, 
a variant of the 
PBP model where 
particles produce a chemical on one hemisphere and consume it with the same rate on the other one (zero net production), allows to explore patterns resulting from 
the Janus instability. 
This variant leads to dynamic clusters with a finite self-limited size, shown in Fig.~\ref{dynclusters_exp} G,H.

\section{\rule{10pt}{10pt} Beyond the Active Brownian Particle Model}

\subsection{Chiral Active Matter}
\begin{figure}
\includegraphics[width=0.6\textwidth]{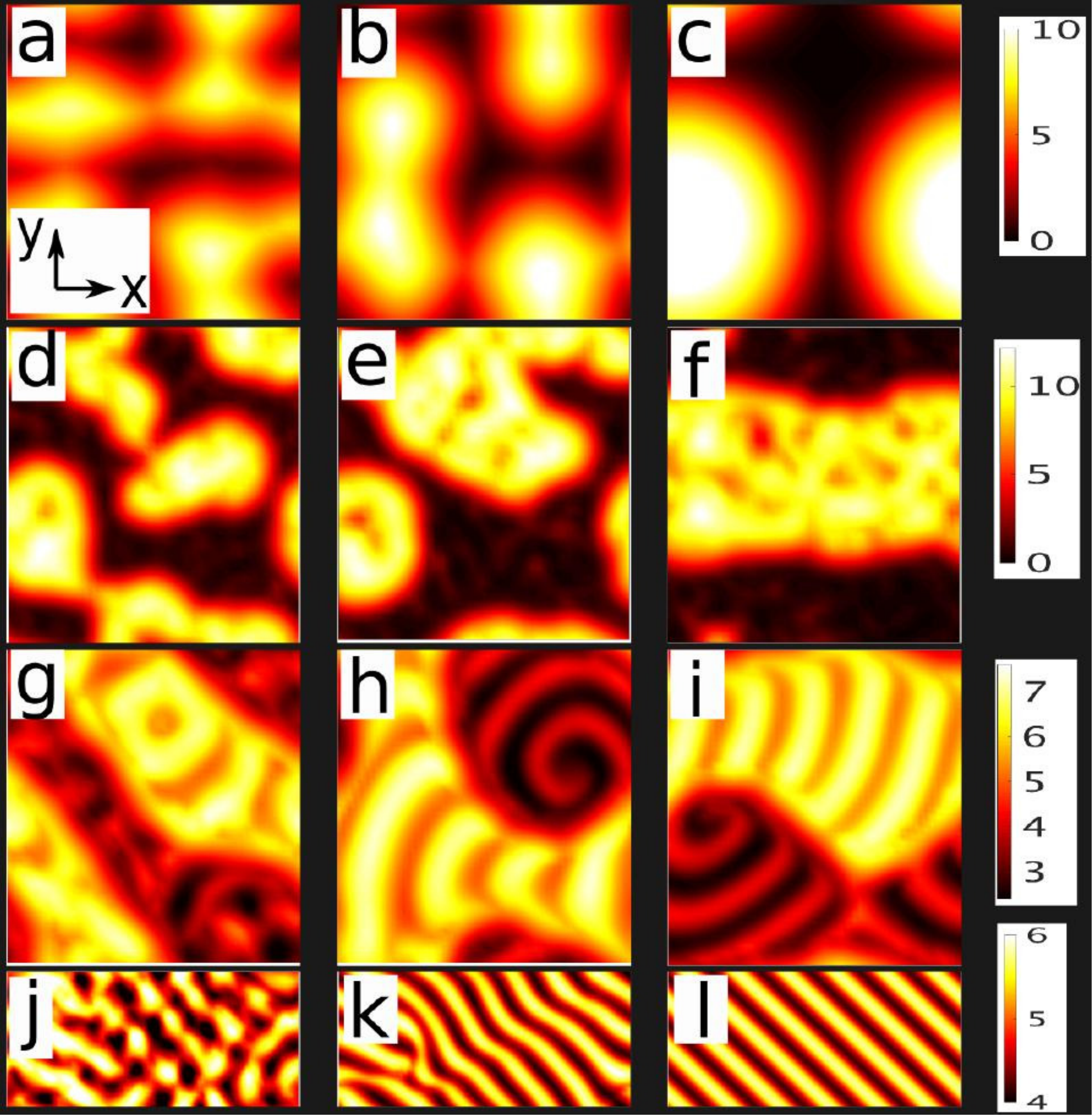}
\caption{Patterns in chemically interacting (signalling) chiral active particles. 
Each row shows the particle density field $\rho({\bf r},t)$ at three consecutive points in time. 
The first row (a-c) shows the particle dynamics essentially following the  
Keller-Segel instability for slow rotations. The second and third row shows phase separation with clusters (d-f) and wave patterns (g-i) forming 
within both the dense and the dilute phase for moderate rotation frequencies.
The bottom row shows global travelling wave patterns (j-l).
Fig. from ref.\cite{Liebchen2016}.} 
\label{figrotors}
\end{figure}

Many active particles swim in circles. Such a chiral motion occurs naturally for asymmetric self-propelled particles 
(e.g. for L-shaped swimmers \cite{Kuemmel2013}). 
Phoretic interactions in such particles introduce a competition between rotations (circular swimming) and 
phoretic alignment up/down gradients of the self-produced chemical field. 
Depending on whether rotations or phoretic alignment dominates (statistically) the polarization density may either rotate or remain phase locked. 
Specifically for particles which synchronize their rotations due to alignment interactions among them, a simplified model \cite{Liebchen2016}
reveals a wealth of possible patterns including travelling wave patterns and spiral patterns (\ref{figrotors}). 
Some parameter regimes even allow for phase separation combined with the emergence of wave patterns within the dense and the dilute phase.
Here, phase separation is based on a locking of the polarization direction of the particles at the interface between the dilute and the dense
phase, such that particles reaching the interface migrate from the dilute to the dense phase (balancing opposing diffusion).
Besides ref.\cite{Liebchen2016}, which focuses on the case where particles are (completely) synchronized, most aspects of the physics of 
chemically interacting chiral active matter remain to be studied. 

\subsection{Further Extensions}
So far, for convenience we have mainly focused on chemotactic alignment. Generally, however chemotactic drift effects should be also important. 
In particular, chemoattractive drifts alone \cite{Liebchen2018PIs}, as well as their combination with alignment (taxis) down the chemical gradient 
can lead to dynamic clustering \cite{Pohl2014,Liebchen2018PIs}. The opposite case
of a repulsive drift and a chemotactic alignment towards (up) the chemical gradient can 
lead to an oscillating state between a collapse and its dissolution \cite{Pohl2015}.
\\While we have focused here on one chemical field representing an effective combination of the fuel chemical and the product species,
the case of two chemical species has been explicitly discussed in \cite{Saha2014}, providing a catalogue of possible instabilities of the uniform phase, which 
can lead e.g. to clustering and collective oscillations. 
\\The case of two different colloidal species producing chemicals 
can lead to nonreciprocal interactions \cite{Ivlev2015} among the colloids \cite{Bartnick2016}, 
which can result in the formation of ``active molecules''
\cite{Soto2014}. These molecules consist of non self-propelled colloids which interact non-reciprocally via self produced (chemical) fields, leading to 
propulsion of the pair. 
Active molecules have recently been demonstrated experimentally in \cite{Niu2017a} based on ion exchange particles interacting via charge-neutral ionic gradients with 
colloids and in 
a light-controllable setup based on colloids \cite{Schmidt2018} leading to
stators, rotators, spinners and linear swimmers \cite{Schmidt2018}.
\\The influence of hydrodynamic interactions on the collective behavior of chemically interacting active particles is still unclear in many cases. However, 
interesting special cases have been discussed in the literature. 
For chemotactic run and tumble bacteria \cite{Lushi2012} it has been shown that hydrodynamic interactions 
can induce 'mixed phase'. For chemoattractive active colloids, in turn, 
hydrodynamic interactions might arrest the chemotactic collapse induced by the Keller-Segel instability \cite{Scagliarini2016}
and for thermorepulsive particles the interplay of hydrodynamics can lead to swarm like structures \cite{Wagner2017}.
\\While showing remarkably close analogies to signalling among cells and microorganisms, research on synthetically signalling active particles may shed new light on 
biological phenomena. One recent example is \cite{Mukherjee2018} applying a variant of the field theory of the PBP model \cite{Liebchen2015,Liebchen2017phoretic} 
to growing bacterial colonies. 

\section{\rule{10pt}{10pt} Conclusions and Outlook}
We have reviewed the collective behavior of active systems showing synthetic chemotaxis with a focus on the Phoretic Brownian Particle  (PBP)
model and corresponding field theories. 
As we have seen, there is now much evidence that synthetic chemotaxis is responsible for the typical collective behavior of autophoretic microswimmers and other active systems like 
camphor and droplet swimmers and can lead to phenomena such as dynamic clustering and the formation of wave patterns at low particle density. 
However, perhaps owed to the complexity of typical active systems, we are still at the beginning of understanding the 
impact of synthetic signalling in detail. 
Open problems include a detailed characterization of the properties of patterns occurring in systems with repulsive chemical interactions using e.g. amplitude equations, 
structure factors etc.
Further important points would be to perform more microscopic simulations involving a description of relevant chemical reactions to estimate the  
evaporation rate of the chemical field ($k_d$) and to better understand phoretic near field effects. 
Further open problems include an understanding of fuel depletion leading to a dependence of the self-propulsion velocity of the particles to the 
density of a chemical field; such fuel depletion effects might not change the onset of structure formation but should become important once a 
phase involving dense components emerges. 
Further important open points include 
boundary and dimensionality effects, as well as a general understanding 
of the interplay of phoretic and hydrodynamic interactions regarding the large scale collective behavior of autophoretic swimmers. 
Short-range repulsions among the colloids, so far included in simulations but not in theoretical calculations based on the PBP model, could be described
e.g. in the framework of dynamical density functional theory \cite{Rex2007,Menzel2016}. Similarly, dynamical density functional theory could also be used to 
describe interactions among the chemicals perhaps leading to interesting phenomena, e.g. if the chemical has a tendency to phase separate. 
Finally, further interesting problems concern the role of 
phoretic interactions in chiral autophoretic swimmers without explicit alignment interactions and
the development of a theory for synthetic signalling in electrophoretic microswimmers based on electrokinetic equations \cite{Brown2017} probably 
leading to a non-linear coupling of colloids to chemical gradients.

\begin{acknowledgement}
H.L. acknowledges support from the DFG within project LO 418|19-1.
We thank F. Hauke for helping us to produce the conspectus-figure. 
\end{acknowledgement}

\section{Biographical Information}
Benno Liebchen has studied physics in Marburg and Berlin, Germany. In 2014 he received his PhD in theoretical physics at the University of Hamburg. 
He then moved to Edinburgh (UK) where he became a postdoc and a Marie 
Curie fellow. Since 2017 he is a postdoc at
the University of D\"usseldorf. He works on soft matter, active soft matter and nonlinear dynamics 
using theoretical descriptions and computer simulations.  

Hartmut L\"owen studied physics at the University of Dortmund,
Germany. In 1987, he received his PhD degree in physics from Dortmund University on phase
transitions in polaron systems. He then moved to the Ludwig-Maximilians-Universit\"at M\"unchen
where he worked on surface melting and classical density functional theory. During a postdoc
stay in Lyon (France) he started to do research on colloidal
dispersions. Since 1995 he is full professor for Theoretical Physics
at the University of D\"usseldorf (Germany). His research concerns
theoretical descriptions and computer simulations of soft matter
systems. 


\bibliography{literature}
\bibliographystyle{unsrt}

\end{document}